\documentstyle[prl,twocolumn,aps,epsf]{revtex}
\begin{document}
\include{psfig}
\twocolumn[
\hsize\textwidth\columnwidth\hsize\csname@twocolumnfalse\endcsname
%\draft

\title{Charged impurity scattering limited low temperature resistivity 
of low density silicon inversion layers}
\author{S. \ Das Sarma and E. H.\ Hwang}
\address{Department of physics, University of Maryland, College Park,
Maryland  20742-4111 } 
\date{\today}
\maketitle

\begin{abstract}

We calculate within the Boltzmann equation approach the charged 
impurity scattering limited low temperature electronic resistivity 
of low density $n$-type inversion layers in Si MOSFET structures. 
We find a rather sharp quantum to classical crossover in the transport 
behavior in the $0 - 5$K temperature range, with the low density, low 
temperature mobility showing a strikingly strong 
non-monotonic temperature 
dependence, which may qualitatively explain the recently observed 
anomalously strong temperature dependent resistivity in low-density, 
high-mobility MOSFETs.


\end{abstract}
\vspace{0.5in}
]
Several recent publications on low temperature resistivity 
measurements\cite{one,two,three,four} in various low 
density two dimensional (2D) systems report the observation of 
an anomalously strong temperature dependence as a function of 
carrier density, which has been interpreted as evidence for a 
zero temperature two dimensional metal-insulator transition 
(2D M-I-T), which is considered to be forbidden in two dimensions 
(at least for a non-interacting 2D system) by the one parameter 
scaling theory of localization \cite{seven}.
A number of theoretical papers\cite{eight,nine,ten}
have appeared in the literature providing 
many possible resolutions of this seemingly unanticipated (but 
apparently ubiquitous) phenomenon. 
In this Letter we propose a possible theoretical explanation for 
(at least a part of ) the observed phenomena. Our explanation 
is quantitative, microscopic, and physically motivated.
Although our theory is quite general and generic (and thus applicable 
to all the systems \cite{one,two,three,four} exhibiting 
the so-called 2D M-I-T), we specifically consider here the electron 
inversion layer in Si MOSFETs, which is both the original system 
in which the 2D M-I-T was first reported \cite{one} and is also the 
most exhaustively experimentally studied \cite{one,two,three} system 
in this context.
It is important to emphasize that, in contrast to much \cite{eight} of the 
existing theoretical work on the subject, our theory does {\it not} address 
the existence (or not) of a {\it zero temperature} 2D M-I-T, but specifically 
addresses the issue of quantitatively understanding the strikingly unusual 
{\it finite temperature} experimental results on the effective ``metallic'' 
side of the transition. 

We first summarize  the key experimental features of the 
2D M-I-T phenomenon (focusing on Si MOSFETs), emphasizing the 
specific aspects addressed in our theory.
Experimentally one finds a ``critical density'' ($n_c$) separating 
an effective ``metallic'' behavior (for density $n_s > n_c$) from 
an effective ``insulating'' behavior ($n_s < n_c$). We concentrate 
entirely on the effective ``metallic'' behavior in this Letter since 
a 2D metal is ``unusual'' according to the conventional 
theory \cite{seven} and a 2D insulator is not. The experimental 
insulating behavior (for $n_s < n_c$) is quite conventional for 
a strongly localized semiconductor and can be understood using 
standard transport models \cite{nine,ten}. 
The effective ``metallic'' behavior is characterized by a strong 
drop in the temperature dependent resistivity, $\rho(T)$, at low 
temperatures ($0.1 K \le T \le 1 - 3K$) and at low densities 
($n_s \ge n_c$).
This novel and dramatically strong temperature dependence of 
$\rho(T)$, where $\rho(T)$ may drop by a factor of $2 - 10$ at low 
electron densities as temperature decreases from 2K to 100 mK, 
is one of the most significant experimental observations we 
qualitatively explain in this Letter. In addition the experimental 
resistivity, $\rho(T,n_s)$, as a function of temperature and electron 
density shows an approximate ``scaling'' behavior $\rho(T,n_s) \simeq 
\rho(T/T_0)$ with $T_0 \equiv T_0(n_s)$ indicating consistency with 
quantum criticality.
Our theoretical results show the same ``scaling'' behavior with our 
calculated $T_0(n_s)$ having very similar density dependence as 
the experimental observation.
There are interesting aspects of the magnetic field and the electric 
field dependence of the observed resistivity, which we do not address 
here, concentrating entirely on the behavior of $\rho(T,n_s)$ in the $n_s 
\ge n_c$ ``metallic'' regime. It is this ``anomalous metallic'' 
behavior (in the sense of a very strong metallic temperature dependence 
of the resistivity in a narrow density range above $n_c$) which has 
created the recent interest in the 2D M-I-T phenomena since 
in general, the 
temperature dependent resistivity of a metal should saturate as it 
enters the low temperature Bloch-Gr\"{u}neisen regime without 
manifesting any strong temperature dependence.

Our theory, which provides good qualitative agreement with the 
existing experimental data on the metallic ($n_s > n_c$) side of the 
transition, is based on two essential assumptions: (1) transport is 
dominated by charged impurity scattering centers (with a density of 
$N_i$ per unit area) which are randomly distributed at the interface; 
(2) the M-I-T at $n_s = n_c$ is characterized by a ``freeze-out'' of 
free carriers due to impurity binding --- the free carrier density 
responsible for ``metallic'' transport is thus ($n_s-n_c$) for 
$n_s > n_c$, and on the insulating side, $n_s < n_c$, the free 
carrier density (at $T=0$) is by definition zero.
Some justifications for these assumptions have been provided in 
ref. \onlinecite{nine} although our current model transcends the 
specific scenario envisioned in ref. \onlinecite{nine} and is
more general. In contrast to ref. \onlinecite{nine}, we do not specify 
any particular 
mechanism for the carrier freeze-out and accept it as an experimental 
fact. We note that we could extend our model and go beyond the above 
two assumptions, for example, by making the effective free carrier 
density $n=(n_s-n_c)\theta(n_s-n_c) + n_a(T)$, where $n_a(T)$ is a 
thermally activated 
contribution to the carrier density 
(this relaxes the second assumption), and/or by 
introducing additional scattering mechanisms such as the short-range 
surface roughness scattering (this relaxes the first assumption). 
These extensions beyond our two essential approximations will 
undoubtedly produce better quantitative agreement between our theory 
and experiment (at the price of having  unknown adjustable 
parameters). We, however, refrain from such a generalized theory, 
because we 
believe that the minimal theory, constrained by our two stringent 
assumptions and thus allowing for only one unknown parameter (the 
charged impurity density $N_i$) which sets the overall scale of 
resistivity in the system, already catches much of the essential 
physics in the problem.

We use the finite temperature Drude-Boltzmann theory to calculate 
the ohmic resistivity of the inversion layer electrons taking only 
into account long range Coulombic scattering by the random static 
charged impurity centers with the electron-impurity Coulomb 
interaction being screened by the 2D electron gas in the random 
phase approximation (RPA). The resistivity is given by $\rho = 
\sigma^{-1}$, where the conductivity $\sigma = ne^2 <\tau>/m$ with 
$m$ as the carrier effective mass, and $<\tau>$ is the energy
averaged finite temperature scattering time:
\begin{equation}
<\tau>=\frac{\int dE E \tau(E) \left ( -\frac{\partial f}{\partial E}
\right )}{\int dE E \left ( - \frac{\partial f}{\partial E} \right )},
\end{equation}
where $f(E)$ is the Fermi distribution function, $f(E) =\{ 1+\exp[ 
(E-\mu)]/k_B T \}^{-1}$ with $\mu(T,n)$ as the finite temperature chemical 
potential of the free carrier system determined self-consistently. 
The energy dependent scattering time $\tau(E)$ for our model of 
randomly distributed interfacial impurity charge centers is given by
\begin{equation}
\frac{1}{\tau(E)} = \frac{2\pi N_i}{\hbar}\int\frac{d^2k'}{(2\pi)^2}
\left |\frac{v(q)}{\varepsilon(q)}\right |^2 (1-\cos\theta) \delta\left (
\epsilon_{\bf k} - \epsilon_{\bf k'} \right ),
\end{equation}
with $q = |{\bf k} - {\bf k}'|$, $\theta \equiv \theta_{{\bf kk}'}$
is the scattering angle between ${\bf k}$ and ${\bf k}'$, 
$E = \epsilon_{\bf k} = \hbar^2k^2/2m$, 
$\epsilon_{\bf k'} = \hbar^2k'^2/2m$, $v(q)$ is the 2D Coulomb 
interaction between an electron and an impurity, and $\varepsilon(q) 
\equiv \varepsilon(q;\mu,T)$ is the 2D finite temperature static RPA 
dielectric (screening) function \cite{eleven,twelve}.
In calculating the Coulomb interaction and the RPA dielectric function 
in Eq. (2) we take into account subband quantization effects in the 
inversion layer through the lowest subband variational wavefunction 
\cite{eleven}.
We note that there are two independent sources of temperature 
dependence in our calculated resistivity --- one source is the energy 
averaging defined in Eq. (1) and the other is the explicit temperature 
dependence of the finite temperature dielectric function $\varepsilon(q)$ 
which produces a direct temperature dependence through screening in 
Eq. (2). At very high temperatures, when $T \gg T_F$ with $T_F \equiv 
\mu(T=0)/k_B$ as the free carrier Fermi temperature, the system
is classical and it is easy to show that 
Eq. (1) leads to a decreasing resistivity with increasing temperature:
$\rho(T) \sim A (T/T_F)^{-1}$ for $T \gg T_F$. In the quantum regime, 
however, energy averaging by itself produces 
a weak quadratic (negative) temperature dependence according to Eq. (1): 
$\rho \sim \rho(T=0) - B(T/T_F)^2$, for $T \ll T_F$. For Si inversion
layer, however, this low temperature negative temperature dependence is 
overwhelmed \cite{eleven,twelve,thirteen} by the temperature dependence 
of the screening function  in Eq. (2) which 
dominates $2k_F$ scattering --- this phenomenon 
arises from the specific form of the 2D screening function 
which is a constant upto $q=2k_F$, and has a cusp at 
$2k_F$ at $T=0$. This strong temperature dependence arising from the 
low temperature screening function produces a linear rise in the 
low temperature ($T \ll T_F$) resistivity with increasing temperature 
according to Eq. (2): $\rho(T) \sim \rho(T=0) + C (T/T_F)$, 
for $T \ll T_F$.
This linear temperature dependence is, however, cut off at very low 
temperatures due to the rounding of the sharp corner in the 2D screening 
function by impurity scattering 
effects \cite{twelve,thirteen,fourteen} --- at very low 
temperature $T \ll T_D$ where $T_D$ ($=\Gamma/\pi k_B$ with $\Gamma$ as 
the collisional broadening) is the collisional broadening induced Dingle 
temperature, the explicit temperature dependence of $\varepsilon(q,T)$ is 
suppressed. At the densities and temperatures of interest 
in the 2D M-I-T phenomena all of these distinct physical effects are 
operational, and the actual behavior of $\rho(T,n)$ could be quite 
complicated because the four different asymptotic mechanisms discussed 
above compete with each other as the system crosses over from a 
non-degenerate classical ($T > T_F$) to a  strongly screened 
degenerate quantum 
($T << T_F$) regime. We note that in general the temperature dependence 
is non-monotonic (particularly at lower densities where the energy 
averaging effects are significant), as has been experimentally 
observed \cite{one,two,three,four}, because the temperature 
dependence of Eq. (1) by itself produces a negative temperature 
coefficient whereas screening through Eq. (2) produces a positive 
temperature coefficient.

In Fig. 1 we show our numerically calculated resistivity for the Si-15 
sample of ref. \onlinecite{one}. 
We show calculated $\rho(T)$ 
as a function of $T$ in Fig. 1 for several values of $n_s > n_c$.
We use several different Dingle temperatures to incorporate
\cite{twelve,thirteen,fourteen} 
the impurity scattering induced 

\begin{figure}
\epsfysize=9.cm
\epsffile{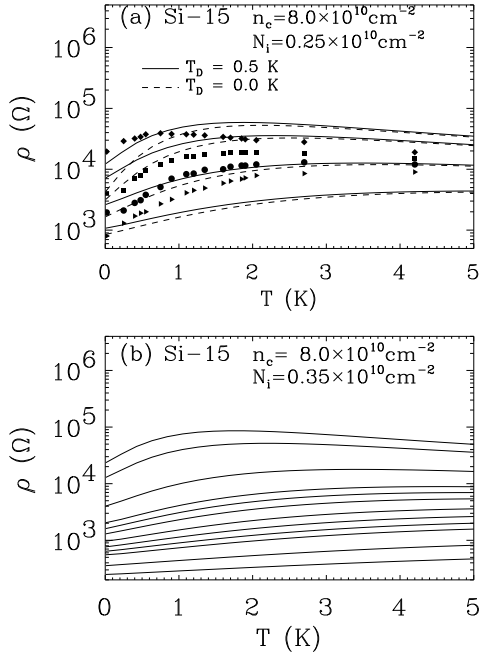}
\caption{
The calculated resistivity $\rho(T)$
for the Si-15 sample of ref. [1]. 
We use (a) the fixed Dingle temperature, 
$T_D=0K$ (dashed lines) and $T_D = 0.5K$ (solid lines), for various electron
densities, $n_s =$ 8.9, 9.2, 10.2, 12.4$\times 10^{10}cm^{-2}$ 
(from top to bottom), and the experimental data 
points are taken from ref. [1];
(b) the density dependent Dingle temperature for 
various densities, $n_s =$ 8.9, 9.2, 10.2, 11.3, 11.8, 12.4, 13.5,
14.5, 15.5, 16.5, 20.0, 24.2$\times 10^{10}cm^{-2}$ (from top to bottom).}
\end{figure}

\noindent
collisional broadening corrections in 
the screening function, including the pure RPA ($T_D=0$) case which 
completely neglects collisional broadening effects on screening.
In Fig. 1(b) we show the calculated Si-15 results where the 
Dingle temperature varies 
as a function of electron 
density. For each density the appropriate 
$T_D$ (going into the screening calculation) is determined from the 
resistivity for that particular density. 
In Fig. 1(b) the temperature dependence of 
$\rho(T)$ at low temperatures is strongest at intermediate densities 
somewhat away from $n_c$ whereas in Fig. 
1(a)  the temperature dependence of $\rho(T)$ becomes stronger 
as one approaches $n_c$, and is the strongest at the lowest density. 
This arises from the competition in screening among $T$, 
$T_F$, and $T_D$ --- at the lowest densities the temperature dependence 
is moderated by having relatively high values of $T_D$ whereas at high 
densities the temperature dependence is suppressed by the large value 
of $T_F$, implying that the strongest temperature dependence of $\rho(T)$ 
occurs at intermediate densities where neither $T_D$ nor $T_F$ is too high. 
Putting $T_D=0$ leads to  stronger temperature dependence 
because the temperature dependence of screening is not cut off at low
temperatures as it is in the $T < T_D < T_F$ regime for the $T_D \neq 0$
results.

The impurity density $N_i$  has been fixed by 
demanding agreement between experiment and theory 
{\it at high temperatures} 
($T=5K$) and the {\it highest densities}. The impurity density $N_i$ 
thus sets 
the scale of the overall resistivity ($\rho \propto N_i$), and does not 
affect the calculated $T$ and $n_s$ dependence of $\rho(T,n_s)$. It is 
important to emphasize that $N_i$ values needed in our calculation to 
obtain quantitative agreement with experiment are in the reasonable
range of $N_i \sim 10^{10} cm^{-2}$, which is known 
\cite{nine,eleven,thirteen} to be the typical effective random 
charged impurity scattering center density in high mobility Si 
MOSFETs. Since $N_i$ is the only ``free'' parameter of our theory, 
it is significant that we obtain a  reasonable value for 
$N_i$  in order to achieve  
agreement between theory and experiment. 
We emphasize that our theory is valid even if the metallic ($n_s > n_c$)
system is actually weakly localized as long as the effective localization 
length is larger than the system size or the phase coherence length.
In general, our calculated 
resistivity is higher (by $25 \sim 40 \%$) than the experimental values 
at low densities ($n_s \geq n_c$), i.e., our theory predicts a somewhat 
stronger $n_s$ dependence of $\rho$ than that observed experimentally. 
This discrepancy can be corrected by adding an activated carrier density 
$n_a(T)$ to our effective carrier density $n = (n_s-n_c) + n_a$, which 
produces the strongest effect at the lowest densities (and essentially 
no effect at higher densities), and would reduce $\rho(T)$ at lower 
densities. One can also use a variable impurity density $N_i(n_s)$ 
which varies with the gate voltage (following the spirit of 
ref. \onlinecite{ten}), and is lower at lower values of $n_s$, 
again producing quantitative 
agreement between theory and experiment.
Given the overall excellent qualitative agreement between our results 
and the experimental data of ref. \onlinecite{one}, we think that these 
refinements of our model are not particularly essential or meaningful.

We show the experimental data points for Si-15 taken from ref. 
\onlinecite{one} in Fig. 1(a) to give an idea about the level of 
agreement between our calculation and the experimental results. 
We do not attach particularly great significance to the 
{\it quantitative} agreement achieved in Fig. 1(a) because of 
the various 
approximations in our theory. We do emphasize, however, that our 
calculations catch all the essential {\it qualitative} features of 
the low temperature experimental data \cite{one,two}. We obtain the 
observed non-monotonicity in $\rho(T)$ at low densities and also the 
strong drop in $\rho(T)$ at low densities in the $0.1 \sim 2 K$ 
temperature range. Consistent with the experimental observations 
our calculated low density $\rho(T)$ could drop by an order of 
magnitude for $1 - 2 K$ change in the temperature. Our high 
density results show weak monotonic increasing $\rho(T)$ with 
increasing $T$ similar to experimental observations 
\cite{one,two,three,four}. We have carried out calculations for 
all the reported Si samples (as well as GaAs samples) in the 
literature \cite{one,two,three,four}, and our level of qualitative 
agreement with experiment in uniformly good (typically as good as it 
is in Fig. 1) for all the existing experiments.
For lower mobility samples, our calculated temperature dependence is

\begin{figure}[t]
\epsfysize=6cm
\epsffile{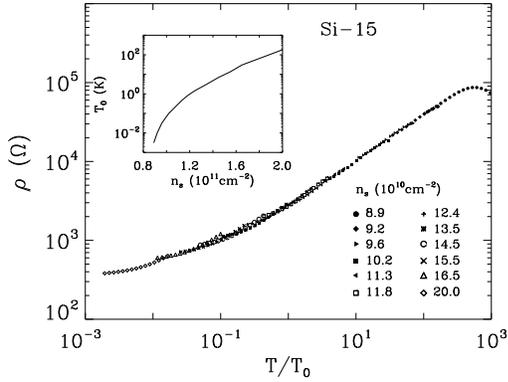}
\caption[]{
The calculated scaling behavior of the resistivity from Fig. 1(b).
Inset shows density dependence of the scaling parameter $T_0$. 
}
\end{figure}

\noindent
weaker (consistent with experimental findings) because screening is 
suppressed by stronger impurity scattering effects (through a higher 
value of level broadening or Dingle temperature).

In Fig. 2 we show our calculated ``scaling'' properties of $\rho(T,n_s) 
\simeq \rho(T/T_0)$ with $T_0 \equiv T_0(n_s)$ for the Si-15 results shown 
in Fig. 1(b). 
The scaling is obtained entirely numerically by 
obtaining the $T_0$ which gives the best scaling fit to the calculated 
$\rho(T,n_s)$. Comparing 
Fig.2 with the corresponding experimental 
scaling plots \cite{one} of resistivity we conclude that 
our theoretical scaling behavior of $\rho(T,n_s)$ is about as good as 
the corresponding experimental scaling. In particular, our $T_0(n_s)$, 
shown as an inset in Fig. 2, agrees reasonably well with the experimental 
results \cite{one}. We obtain very similar ``scaling'' 
results for the 
other Si samples in refs. \onlinecite{one,two}.
The ``scaling'' we obtain in Fig. 2 underscores the important point that 
the experimentally observed scaling behavior in a narrow ($T,n_s$) range 
does not necessarily imply quantum criticality.

Before concluding we point out the approximations made in our calculations.
We have assumed uncritically that the Drude-Boltzmann transport theory, 
which is extensively and successfully employed \cite{eleven} in the 
device simulation of Si MOSFETs, applies to the problem being 
studied. Our main justification for applying the standard transport 
theory to the current problem is our belief that such a ``zeroth order'',
``one-parameter'' ($N_i$ being the only parameter in our model) theory 
must be applied to the problem and compared with the experimental data 
before one can discuss more speculative (and calculationally difficult) 
approaches\cite{eight}. The fact that such a zeroth order theory 
already obtains good qualitative agreement with the experimental 
results indicates that charged impurity scattering, carrier binding 
and freeze-out, temperature and density dependence of 2D screening, 
and classical to quantum crossover (in the $T=0 - 5K$ range) are playing 
significant roles in the experiments and cannot be neglected in any 
theoretical analysis of the ``2D M-I-T'' phenomenon. Our other 
approximations of using the RPA screening (we actually incorporate a 
2D Hubbard local field correction\cite{fifteen} in our screening, 
which has no 
qualitative effect on our results) and the Dingle temperature 
approximation to incorporate collisional broadening effects on screening
are quite reasonable (at least qualitatively) within our model and 
approximation scheme, and may be systematically improved (with a 
great deal of work) if future experiments warrant such a quantitative 
improvement of the theory.
It is important to emphasize that quantum corrections, including 
localization effects, are left out of our semi-classical 
Drude-Boltzmann theory. In providing some justification we mention 
that the effective dimensionless parameter $k_F l$ (where $l$ is the 
transport mean free path) is typically 2 or
larger in our results (at $T = 0$, which we believe to be the 
appropriate limit to consider), and therefore a Boltzmann theory may 
have reasonable validity. We estimate weak localization effects to 
be substantially weaker than the temperature dependence shown in Fig. 1 
in the experimental temperature range ($T > 50 mK$) in refs. 
\onlinecite{one,two}. 
The fact that the observed temperature dependence, particularly at
lower temperatures, is somewhat stronger than our calculated results
may very well be the manifestation of quantum fluctuation or interaction 
effects neglected in our theory.
An important approximation of our theory (consistent with the Drude-Boltzmann 
approach) is the neglect of inelastic electron-electron interaction, which 
may well be significant in the low density 2D systems of experimental 
relevance. For example, it is possible that the insulating system 
($n_s < n_c$) is an electron glass (arising from the competition/frustration 
between interaction and disorder). While a quantitative theory including 
disorder and interaction effects is extremely difficult, we speculate that 
our Boltzmann theory (in particular the quantum-classical crossover which 
leads to the strong temperature dependence) is sufficiently robust so that 
our qualitative conclusions will remain unaffected.

We conclude by emphasizing the specific salient features of the 
qualitative agreement between our theory and experimental 
data on the metallic side: (1) strong temperature dependence at low 
and intermediate densities ($n_s \geq n_c$);
(2) non-monotonicity in $\rho(T)$, arising from quantum-classical crossover,
at low values of $n_s \geq n_c$ 
where $\rho(T)$ increases weakly with decreasing $T$ at higher 
temperatures and decreases strongly with $T$ at lower temperatures; 
(3) scaling of $\rho(T,n_s) \simeq \rho(T/T_0)$ with the theoretical 
$T_0(n_s)$ agreeing with the experimental results; (4) our calculated 
zero temperature conductivity, $\sigma(T=0,n_s) = 1/\rho
(T\rightarrow 0, n_s)$, shows an approximately (within $25 \%$)
linear density dependence,
$\sigma(T=0) \propto n = (n_s - n_c)$, which is consistent 
with the $T \rightarrow 0 $ extrapolation of the experimental 
\cite{one,two} resistivity
and also with several other experimental \cite{sixteen} findings
[this dependence, $\sigma(T=0) \propto (n_s - n_c)$, also 
supports our basic freeze-out or binding model]; (5) for increasing 
disorder (i.e., for lower mobility samples) we predict an increasing 
$n_c$ and weaker temperature dependence, as observed experimentally.

%\section*{ACKNOWLEDGMENTS}
This work is supported by the U.S.-ARO and the U.S.-ONR.

\end{document}